%% file: ICMLA_2021.tex
\documentclass[conference]{IEEEtran}
\IEEEoverridecommandlockouts
\usepackage[noadjust]{cite}
\usepackage{amsmath,amssymb,amsfonts}
\usepackage{algorithmic}
\usepackage{graphicx}
\usepackage{textcomp}
\usepackage{xcolor}
\usepackage{multicol}
\usepackage{float}
\def\BibTeX{{\rm B\kern-.05em{\sc i\kern-.025em b}\kern-.08em
    T\kern-.1667em\lower.7ex\hbox{E}\kern-.125emX}}
\graphicspath{{./figures/}{../figures}}

\begin{document}

\title{Data-driven and Automatic Surface Texture Analysis Using Persistent Homology\\ 
%{\footnotesize \textsuperscript{*}Note: Sub-titles are not captured in Xplore and
%should not be used}
\thanks{This material is based upon work supported by the National Science Foundation under Grant Nos. CMMI1759823 and DMS-1759824 with PI FAK.
\vspace{\baselineskip}

\noindent © 2021 IEEE.  Personal use of this material is permitted.  Permission from IEEE must be obtained for all other uses, in any current or future media, including reprinting/republishing this material for advertising or promotional purposes, creating new collective works, for resale or redistribution to servers or lists, or reuse of any copyrighted component of this work in other works.

}
}

\author{\IEEEauthorblockN{Melih C. Yesilli}
\IEEEauthorblockA{\textit{Department of Mechanical Engineering} \\
\textit{Michigan State University}\\
East Lansing, Michigan \\
yesillim@egr.msu.edu}
\and
\IEEEauthorblockN{Firas A. Khasawneh}
\IEEEauthorblockA{\textit{Department of Mechanical Engineering} \\
\textit{Michigan State University}\\
East Lansing, Michigan \\
khasawn3@egr.msu.edu}
}

\maketitle

\begin{abstract}
Surface roughness plays an important role in analyzing engineering surfaces. It quantifies the surface topography and can be used to determine whether the resulting surface finish is acceptable or not. 
Nevertheless, while several existing tools and standards are available for computing surface roughness, these methods rely heavily on user input thus slowing down the analysis and increasing manufacturing costs.  
Therefore, fast and automatic determination of the roughness level is essential to avoid costs resulting from surfaces with unacceptable finish, and user-intensive analysis.
In this study, we propose a Topological Data Analysis (TDA) based approach to classify the roughness level of synthetic surfaces using both their areal images and profiles. We utilize persistent homology from TDA to generate persistence diagrams that encapsulate information on the shape of the surface. We then obtain feature matrices for each surface or profile using Carlsson coordinates, persistence images, and template functions. We compare our results to two widely used methods in the literature: Fast Fourier Transform (FFT) and Gaussian filtering. The results show that our approach yields mean accuracies as high as $97$\%. We also show that, in contrast to existing surface analysis tools, our TDA-based approach is fully automatable and provides adaptive feature extraction.
\end{abstract}

\input{sections/sec-intro}

\input{sections/sec-simulation}

\input{sections/sec-methodology}
\input{sections/sec-results_and_discussion}

\bibliography{ICMLA_2021}
\end{document}

%% file: sections/sec-intro.tex
%!TEX root = ../ICMLA_2021.tex
%*****************************
\section{Introduction}
\label{sec:intro}
Surface texture analysis is a prominent field of research with many applications including tribology~\cite{Suh2003}, metrology, remote sensing~\cite{Makarenko2016}, medical imaging ~\cite{Taha2015}, and the marine industry~\cite{Wu2018}. 
One specific active area of research is the fast and automatic feature extraction from image data that reduces the need for the input of expert users. 
In addition to the need for reliable, automatic feature extraction other challenges in surface texture analysis include the size of the data which significantly increases with increasing the resolution. 
% One of the challenges is that size of the data set can increase for small regions of the surfaces with respect to the selected resolution, so computational time also increases.
% Another challenge is that current methods in the literature require require manual inspection of the data set to select parameters.
Therefore, there is a need for adaptive and automatic tools for feature extraction from surface images.

The majority of the tools proposed for roughness analysis of engineering surfaces are based on decomposing the image data using a set of basis functions that can be grouped into three main components: form, waviness and roughness. 
Form contains the lowest frequencies, while waviness is composed of sinosoidal waves in the middle frequency range. 
Larger frequencies are included in the roughness component.
Generally, the majority of surface analysis tools are focused on finding the reference surface or profile.
Depending on the feature extraction tool used, reference surface (profile) is composed of form or it is the combination of both form and waviness. 
The surface roughness can then be obtained by subtracting the form and the waviness from the original surface. 
% The resulting component is analyzed to determine the roughness level of the corresponding surface.

For surface profile analysis, Gaussian filter is one of the most commonly used filters in the literature~\cite{ISO16610,Raja2002,Gurau2004}.
It is used as low-pass filter to obtain a smoother surface, and roughness profile is then obtained by subtracting the filtered profile from the original one. 
Fast Fourier Transform (FFT) is also another widely adopted filtering approach for feature extraction from 1D signals \cite{Yesilli2020} including surface profiles. 
% Yesilli and Khasawneh used the peaks of FFT and Power Spectral Density (PSD) and Autocorrelation (ACF) to detect chatter in turning process~\cite{Yesilli2020}.
For example, Raja and Radhakrishnan used FFT to denoise 1D surface data and obtain the corresponding roughness profiles~\cite{Raja1979}. 
For surface areal data, two dimensional implementations of FFT and Gaussian filter can be utilized \cite{Indahl1998,Peng1997}.
Peng and Kirk applied 2D-FFT to surface images of three different wear particles and used spectral intensity values in angular and radial spectra to identify the type of wear particle \cite{Peng1997}. 
Other decomposition-based signal processing tools such as discrete wavelet transform (DWT) \cite{Raja2002,Goic2016} and discrete cosine transform (DCT) \cite{Goic2016,Lecompte2010} are utilized in surface texture analysis.
DCT and DTW require selecting a threshold to separate form, waviness and roughness components. 
However, there is not any guide on how to select these thresholds in the literature.
Most of the studies which use these approaches work with a small number of data sets and they select the thresholds by trial and error. 
However, this makes the featurization process extremely cumbersome as the size of the data set increases. %Therefore, these methods have low potential of automation.

The limitations in the current methods for surface roughness analysis show us that there is a need for automated and adaptive feature extraction methods. In this study, we propose to use persistent homology which is a tool from Topological Data Analysis (TDA) for quantifying the roughness of surfaces.
%Topological Data Analysis has been used in various application such as chatter detection \cite{Yesilli2019, Yesilli2019a}, time series analysis \cite{Myers2019}, dynamic state detection \cite{Myers2019a} and financial data analysis\cite{Gidea2017}.
Specifically, we use 0D and 1D sublevel set persistence on surface profiles and surface images to compute the sublevel set persistence diagrams.
Then, we utilize Carlsson Coordinates \cite{Adcock2016,Khasawneh2018}, persistence images~\cite{Adams2017}, and template functions~\cite{Perea2019} to extract features from these diagrams.
We use the TDA-based approach on synthetic data sets to identify the level of roughness, and we compare its performance to features extracted using traditional image analysis. 
Our results show that our TDA-based approach can match or outperform the traditional signal processing tools. However, in contrast to traditional methods, all the steps in the TDA approach are automatically performed and there is no need for manual preprocessing. 

This paper is organized as follows. Section~\ref{sec:Simulation} explains how the synthetic data set was obtained. Section~\ref{sec:method} describes traditional, widely used methods in the literature as well as our TDA-based approach. Section~\ref{sec:Results} compares the results from our approach to its traditional counterparts, and discusses the classification accuracies.

\subsection{Our Contribution}
Previous studies on surface roughness analysis utilize traditional signal processing tools to decompose surfaces into their frequency components. 
Specifically,  a threshold value is selected to separate the roughness component from the original surface for DCT and DTW approaches \cite{Goic2016,Lecompte2010}. 
However, there are no standards that guide the selection of these thresholds, thus hindering the possibility of automation. 
In this paper, we eliminate both the manual parameter selection and the decomposition phase for surface roughness analysis, and we provide a fully automated pipeline to analyze engineering surfaces.
Both 3D surfaces and 2D profiles are analyzed using their topological structure, and information gained from this analysis is summarized in persistence diagrams. 
We utilize several techniques to vectorize the resulting persistence diagrams to generate feature matrices.
Another contribution of this work is presenting a new method for quantifying similarity of surfaces. 
Specifically, previous studies focus on the analysis of several surfaces and they do classify the resulting surfaces based on their similarity.   
In contrast, we classify the surfaces with respect to their roughness characteristics thus presenting a new approach for measuring similarities of surfaces since the same features can be used in unsupervised clustering algorithms.

%% file: sections/sec-simulation.tex
%!TEX root = ../ICMLA_2021.tex
%*****************************
\section{Simulation}
\label{sec:Simulation}
\begin{figure}[hbt!]
\centering
\includegraphics[width=0.4\textwidth,keepaspectratio]{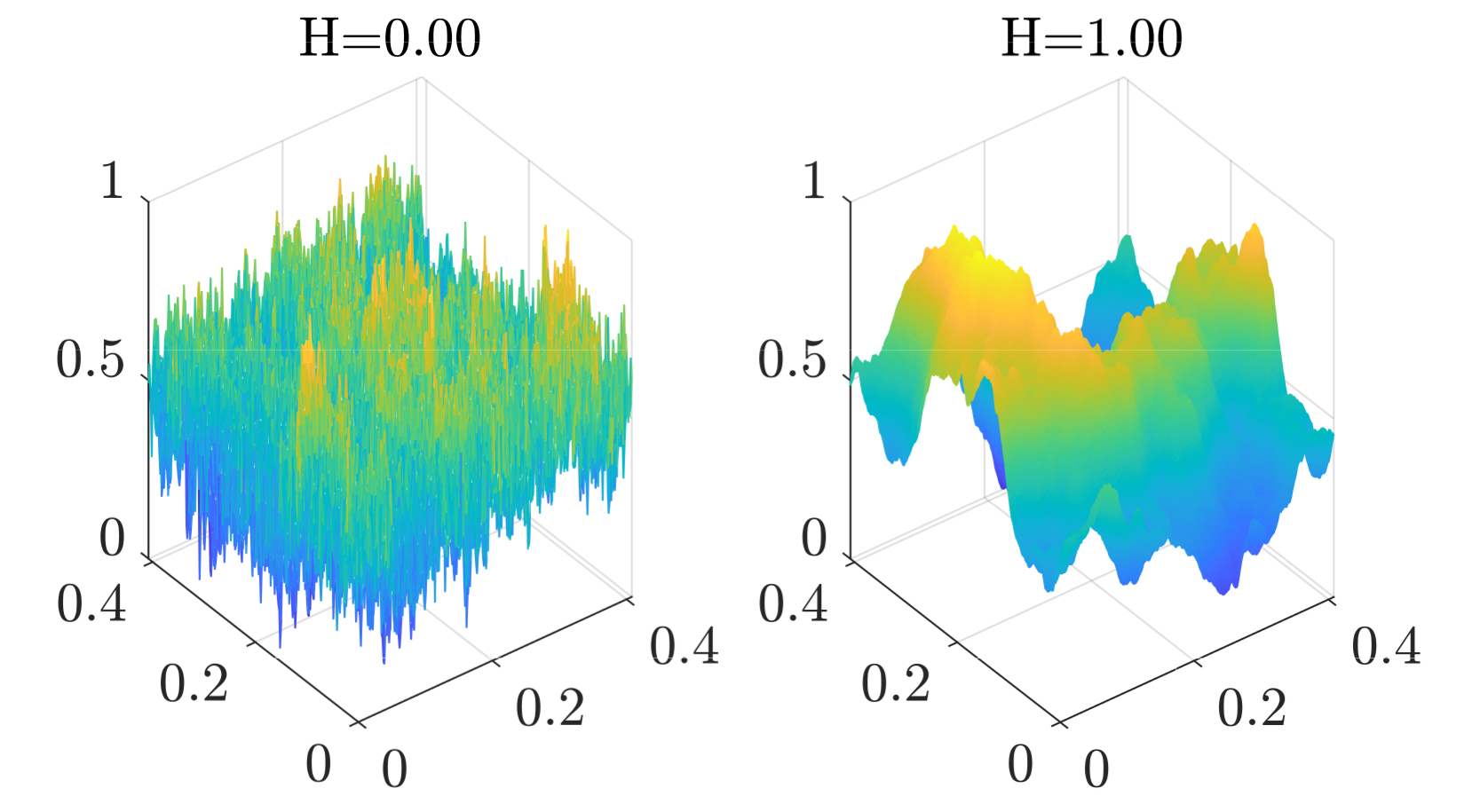}
\caption{The roughest and smoothest surface in the synthetic data set obtained with $H=0$ and $H=1$, respectively.}
\label{fig:example_surfaces}
\end{figure}

We use synthetic surfaces to test the proposed approach, which are generated using the model provided in Ref.~\cite{Mueser2017}. 
The surface roughness of the resulting surfaces is controlled by Hurst roughness parameter $H \in [0, 1]$. As the value of $H$ varies from $0$ to $1$, the generated surface gets smoother and smoother, see the example surfaces in Fig.~\ref{fig:example_surfaces}.

We divided the $[0,1]$ range into 200 intervals and obtained 201 roughness parameters that were then used to generate synthetic surfaces. 
Then, we categorized the resulting surfaces according to their roughness parameter value into three classes. 
The first and last 67 surfaces are categorized as rough and smooth surfaces, respectively. The surfaces in between these two cases were tagged as somewhat rough.

In addition to the generated surface data, we work with surface profiles in this study. 
We extracted six surface profiles in two perpendicular directions of surfaces.
Therefore, we have totally 1206 surface profiles whose labels match the underlying generated surfaces.

%% file: sections/sec-methodology.tex
%!TEX root = ../ICMLA_2021.tex
%*****************************
\section{Methodology}
\label{sec:method}
In this section, we briefly explain the feature extraction methods from both surfaces and surface profiles. We categorize the methods we use into two groups: 1) traditional image/signal processing methods and 2) TDA based approach.
For the first one, the general idea is to find a reference surface or a profile and subtract it from the original measurement to obtain the roughness surface or profile. 
Then, we compute height parameters, spatial parameters and hybrid parameters provided in Secs.~4.1-4.3 of Ref.~\cite{ISO21920} for roughness profiles. 
While working with roughness surfaces, height and hybrid parameters are used as features, and they are provided in Secs.~4.2 and 4.4 of Ref.~\cite{ISO25178}.
For 1D peak selection method of FFT, we use the coordinates of the peaks of FFT and PSD plots. 
The angular spectral densities are used as features in the case of the two dimensional FFT.

For the TDA based approach, we use three featurization techniques to generate feature vectors for persistence diagrams: Carlsson Coordinates \cite{Adcock2016,Khasawneh2018}, persistence images\cite{Adams2017}, and template functions\cite{Perea2019}.

\subsection{Gaussian Filtering}
\label{sec:Gaussian}
\subsubsection{\textbf{1D-Implementation}}
Gaussian filtering is one of the most commonly used tools for profile filtering \cite{Raja2002}. 
We implement Gaussian filtering in 1D and 2D to analyze surface profiles and areas, respectively.
The 1D kernel definition is given as \cite{ISO16610},
\begin{equation*}
G(x) = \frac{1}{\alpha \lambda_{c}}\exp\Bigg(-\pi\Big(\frac{x}{\alpha \lambda_{c}}\Big)^{2}\Bigg),
\end{equation*}
where $\alpha = \sqrt{ln2/\pi}$, and $\lambda_{c}$ is the roughness long wavelet cutoff \cite{Raja2002}. 
Cutoff selection is performed with respect to the iterative procedure provided in Ref.~\cite{ASMEB46}, which we summarize below. 
First, we estimate the surface roughness parameter, $R_{a}$ for surface profiles using the expression \cite{ISO21920},
\begin{equation*}
R_{a} = \frac{1}{L}\int_{L} \mid z(x)\mid dx,
\end{equation*}
where $L$ represents the measurement length of the profile.
A cutoff value is chosen from Table 3-3.20.2-1 provided in Ref.~\cite{ASMEB46}. 
Then, we measure $R_{a}$ for the roughness profile after applying the filter using the chosen cutoff value.
If the new $R_{a}$ is outside of the range of the old $R_{a}$, we select a new cutoff with respect to new $R_{a}$. However, if it is larger than the measurement length, the algorithm automatically picks the first chosen value as cutoff. This procedure is for nonperiodic profiles, and one can refer to \cite{ASMEB46} for more details.

After setting the cuttoff value and applying the Gaussian filter, we obtained a filtered profile which is also called roughness mean line. 
Roughness profile is obtained by subtracting the mean line from the original surface profile. 
Then, we compute the profile features provided in Ref.~\cite{ISO21920} to generate the feature matrix for supervised classification.

\subsubsection{\textbf{2D-Implementation}}
The 2D Gaussian kernel expression is given as
\begin{equation*}
G(x,y) = \frac{1}{\sqrt{2\pi \sigma^{2}}}\exp\Bigg(\frac{-x^{2}-y^{2}}{2\sigma^{2}}\Bigg),
\end{equation*}
where $\sigma$ is the standard deviation.
After we compute the kernel in 2D, we convolve the surface measurement with the kernel to obtain the filtered surface.
The convolution is performed using 
\begin{equation*}
I[i,j] = \sum_{u=-W}^{W}\sum_{v=-W}^{W}G[u,v]f[i-u,j-v],
\end{equation*}
where $2\times W+1$ equals the kernel size $K$, $f$ is the surface measurement, and $I$ is the filtered surface. The standard deviation $\sigma$ is defined using the expression, $\sigma = K/6$.
\begin{figure}[t!]
\centering
\includegraphics[width=0.5\textwidth,keepaspectratio]{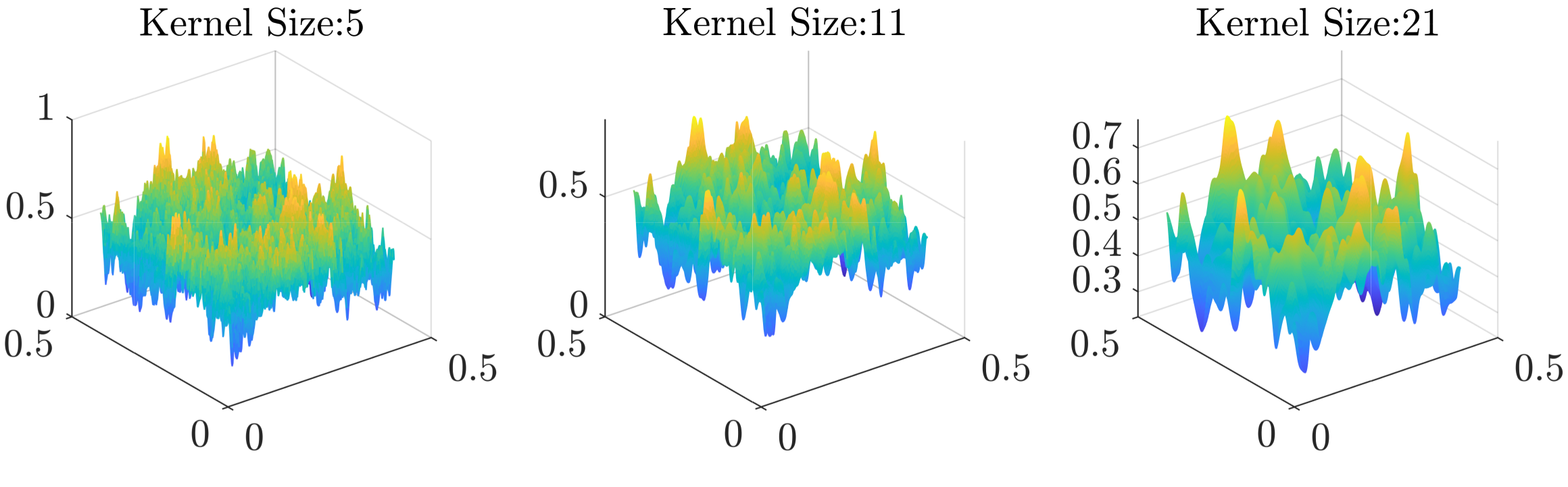}
\caption{Filtered surfaces obtained using kernel sizes 5, 11 and 21.}
\label{fig:Gaussian_kernel_sizes}
\end{figure}

We applied Gaussian filtering in 2D to the roughest surface in the synthetic data set with three different kernel sizes.
The resulting surfaces are provided in Fig.~\ref{fig:Gaussian_kernel_sizes}. It is seen that larger kernel sizes provide smoother filtered surfaces. 
The roughness surface is obtained by subtracting the filtered surface from the original surface. 
Smoother filtered surfaces allows us to have higher frequency components in the roughness surface. 
Therefore, we select a kernel size of 21 and keep it constant in all filtering operations.
Then, we compute areal parameters obtained from Ref.~\cite{ISO25178} on roughness surfaces obtained after filtering. These parameters constitute our features for supervised classification.

\subsection{Fast Fourier Transform (FFT)}
\label{sec:FFT}
\subsubsection{\textbf{1D - Denoising}}
Fast Fourier Transform is one of the mostly adopted signal and image processing tools. 
It was employed to analyze surface profiles in Ref.~\cite{Raja1979}. 
The main idea is to manipulate the spectrum and then apply inverse FFT to obtain a filtered profile.
We applied FFT on the surface profiles, obtained their normalized spectra, and selected a cutoff value between zero and one.  
The amplitudes below that cutoff are set to zero thus eliminating the corresponding frequencies from the data. 
Inverse FFT is then applied to the modified spectrum to yield a mean line profile.
Subtracting the filtered profile from the original one gives us the roughness profile.
\begin{figure}[hbt!]
\centering
\includegraphics[width=0.45\textwidth,keepaspectratio]{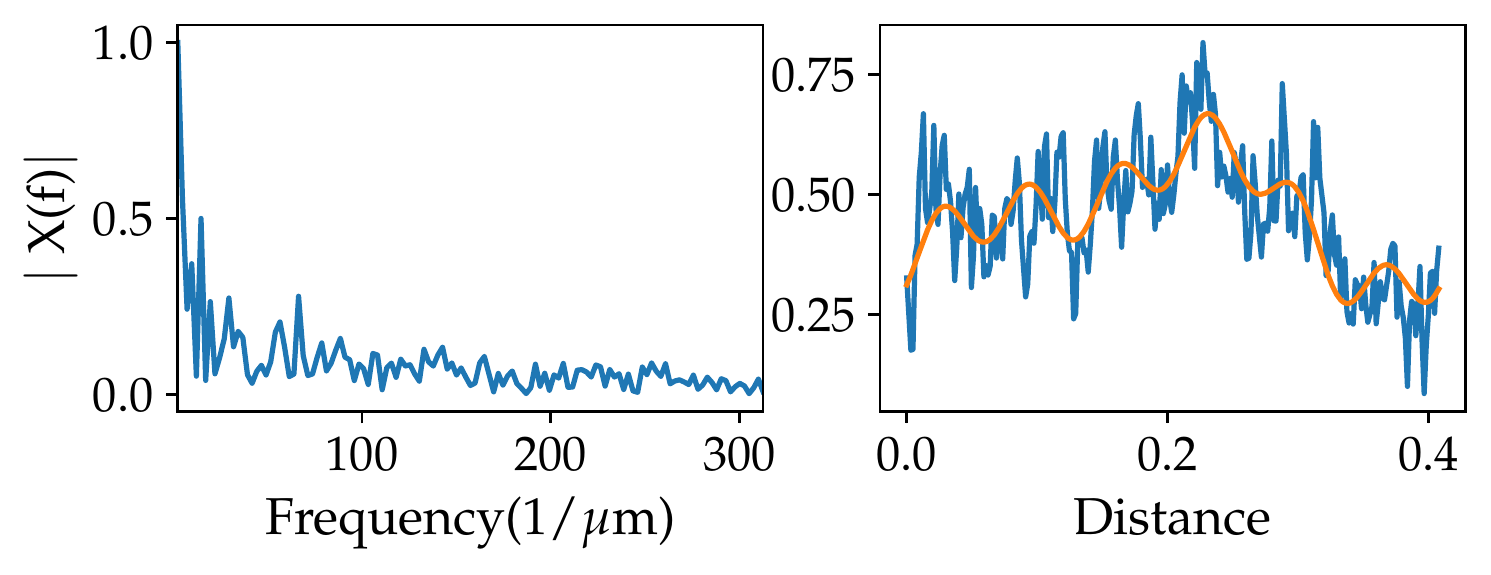}
\caption{(Left) Spectrum of the plots. Filtered profiles obtained from two cutoff values, 0.2 (middle) and 0.4 (right).}
\label{fig:Cutoff_effect}
\end{figure}

An example of a filtered profile is provided in Fig.~\ref{fig:Cutoff_effect}. 
the figure shows that larger cutoff values provide smoother profiles.
Therefore, we choose cutoff value 0.4 to eliminate high frequencies in the filtered profile. Profile parameters are computed for each roughness profile, and a feature matrix is generated.
\subsubsection{\textbf{1D - Peak Selection}}
\begin{figure}[hbt!]
\centering
\includegraphics[width=0.45\textwidth,keepaspectratio]{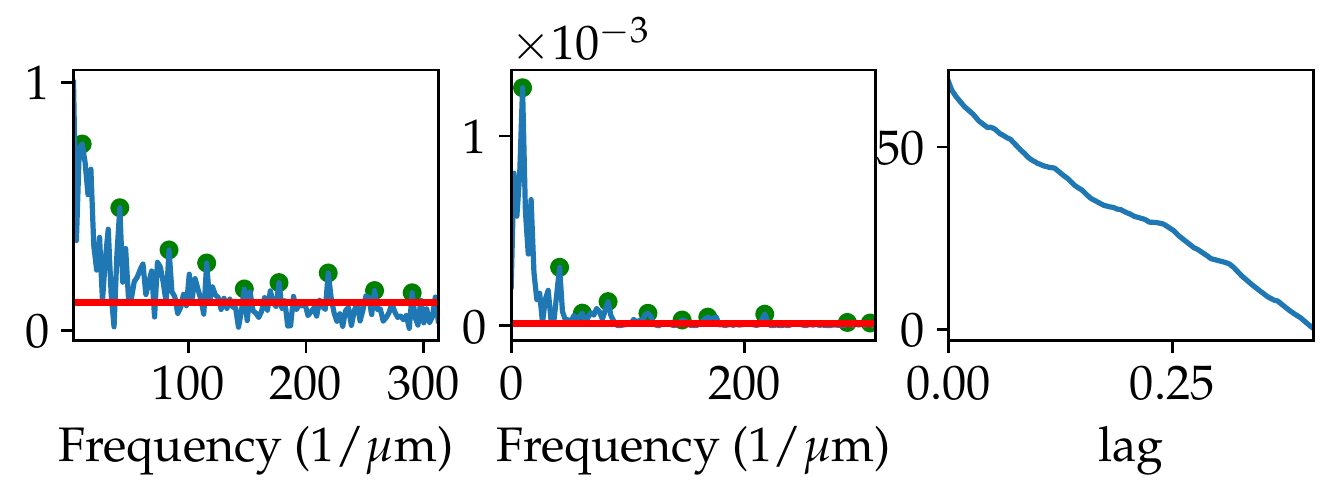}
\caption{Selected peaks for FFT and PSD plots with respect to MPH and chosen MPD values. Red horizontal lines represent the MPH.}
\label{fig:Peaks_FFT_PSD}
\end{figure}

The peaks' coordinates in Fast Fourier Transform, Power Spectral Density (PSD) and Autocorrelation (ACF) plots can be used as features in 1D signals \cite{Yesilli2020}, and that is the approach we implement here for identifying the level of roughness in the synthetic data set.
However, we exclude ACF plots since no peaks are detected in the ACF plots (see Fig.~\ref{fig:Peaks_FFT_PSD}).

First, we compute the FFT and the PSD spectra from the surface profiles.
Then, peak selection is performed with respect to two restriction parameters to locate the true peaks of the spectrum.
These parameters are minimum peak height (MPH) and minimum peak distance (MPD).
MPD is the minimum sample number between two consecutive peaks. 
We selected MPD as 7 and 10 for PSD and FFT plots, respectively. 
The expression for MPH is 
\begin{equation*}
MPH=y_{\rm min}+\alpha(y_{\rm max}-y_{\rm min}),
\end{equation*}
where $y_{\rm min}$ and $y_{\rm max}$ are $40$th and $50$th percentile of the amplitudes in the spectrum, respectively, and we set $\alpha=0.5$.

MPD and the parameters in the MPH expression can be adjusted depending on the data set by \textit{visually} inspecting the selected peaks.
We perform this parameter tuning for three different surface profiles obtained from the roughest surface in the data set.
After several adjustments, we obtained some meaningful peaks for both spectra as shown with an example in Fig.~\ref{fig:Peaks_FFT_PSD}. 
Since manually inspecting all the spectra is time consuming, this parameter tuning for MPD and MPH is performed for only three profiles and the tuned parameters are fixed for all the other profiles.
% It is not guaranteed that the selected parameters will locate true peaks for every data set in the collection, thus this can cause misclassification errors. 
After selecting the peaks, we use their coordinates as our features for classification. 
The user can control the size of the feature matrix by specifying the number of peaks.

\subsubsection{\textbf{2D - Implementation}}
FFT can also be applied to images. 
We apply two dimensional FFT to gray scale synthetic surfaces. 
Areal power spectral density is computed with respect to the formula \cite{Peng1997}
\begin{equation*}
G(n/NT_{x},m/MT_{y}) = \frac{1}{MNT_{x}T_{y}}\mid H(n/NT_{x},m/MT_{y})\mid^{2},
\end{equation*}
where $n = 0,1,\ldots,N-1$ and $m = 0,1,\ldots,M-1$.
$M$ and $N$ are the size of the image, while $T_{x}$ and $T_{y}$ are the sampling intervals in $x$ and $y$ directions. $H(n/NT_{x},m/MT_{y})$ is the 2D Discrete Fourier Transform obtained by using
\begin{equation}
    \resizebox{0.5\textwidth}{!}{%
        $H(n/NT_{x},m/MT_{y}) = \sum_{q=0}^{M-1}\sum_{p=0}^{N-1}h(pT_{x},qT_{y})e^{-j2\pi np/N}e^{-j2\pi mq/M}$%      
        }
\end{equation}
where $h(pT_{x},qT_{y})$ represents the surface measurement, $p = 0,1,\ldots,N-1$ and $q = 0,1,\ldots,M-1$.

\begin{figure}[hbt!]
\centering
\includegraphics[width=0.5\textwidth,keepaspectratio]{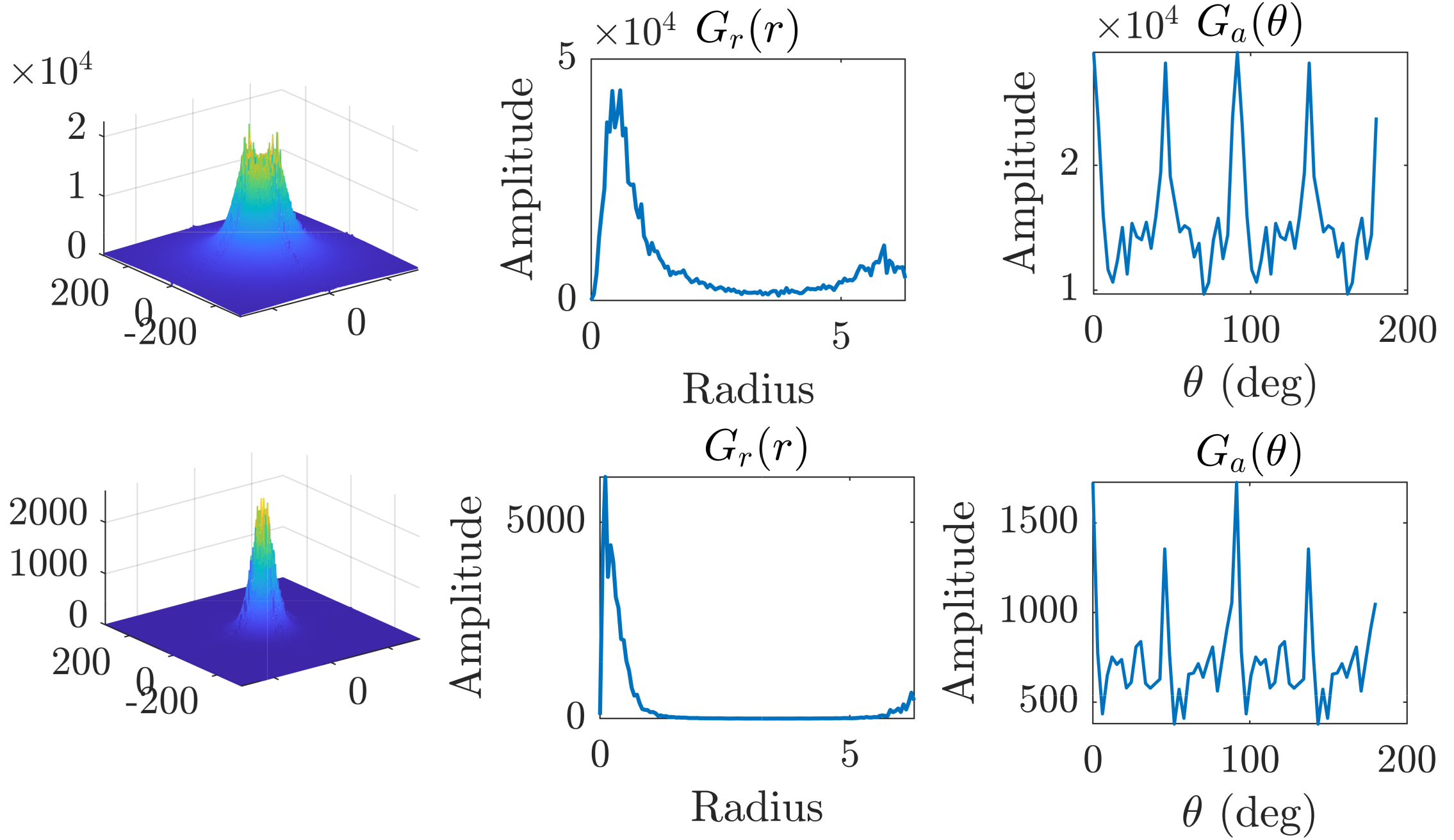}
\caption{APSD plots, radial and angular spectrum of roughest (first row, $H=0$) and smoothest (second row, $H=1$) surfaces. APSDs are obtained after applying the 2D FFT on roughness surfaces.}
\label{fig:APSD_Radial_Angular}
\end{figure}

Areal power spectral density (APSD) plots are analyzed using polar coordinates. 
% Peng and Kirk used angular spectral intensities to identify the type of wear particles \cite{Peng1997}.
In this study, we compute Polar FFT \cite{Averbuch2006} to obtain angular and radial spectrums, similar to \cite{Peng1997}.
In addition, Dong and Stout applied 2D FFT directly to the roughness surface obtained after subtracting the reference surface from the original measurement. 
In this study, we also employ this approach and use the Gaussian filtering explained in Sec.~\ref{sec:Gaussian}.
APSD plots, as well as radial and angular spectra for two surfaces are provided in Fig.~\ref{fig:APSD_Radial_Angular}.
Since there are fewer peaks in the radial spectra, we only take into account the angular spectra.
We use density values of the five peaks in the angular spectrum as features in addition to $\zeta_{\rm max}^{c}$ and $\zeta_{\rm max}^{d}$, given in Ref.~\cite{Peng1997}.

\subsection{Topological Data Analysis (TDA)}
\label{sec:TDA-based}
\begin{figure}[b!]
\centering
\includegraphics[width=0.45\textwidth,keepaspectratio]{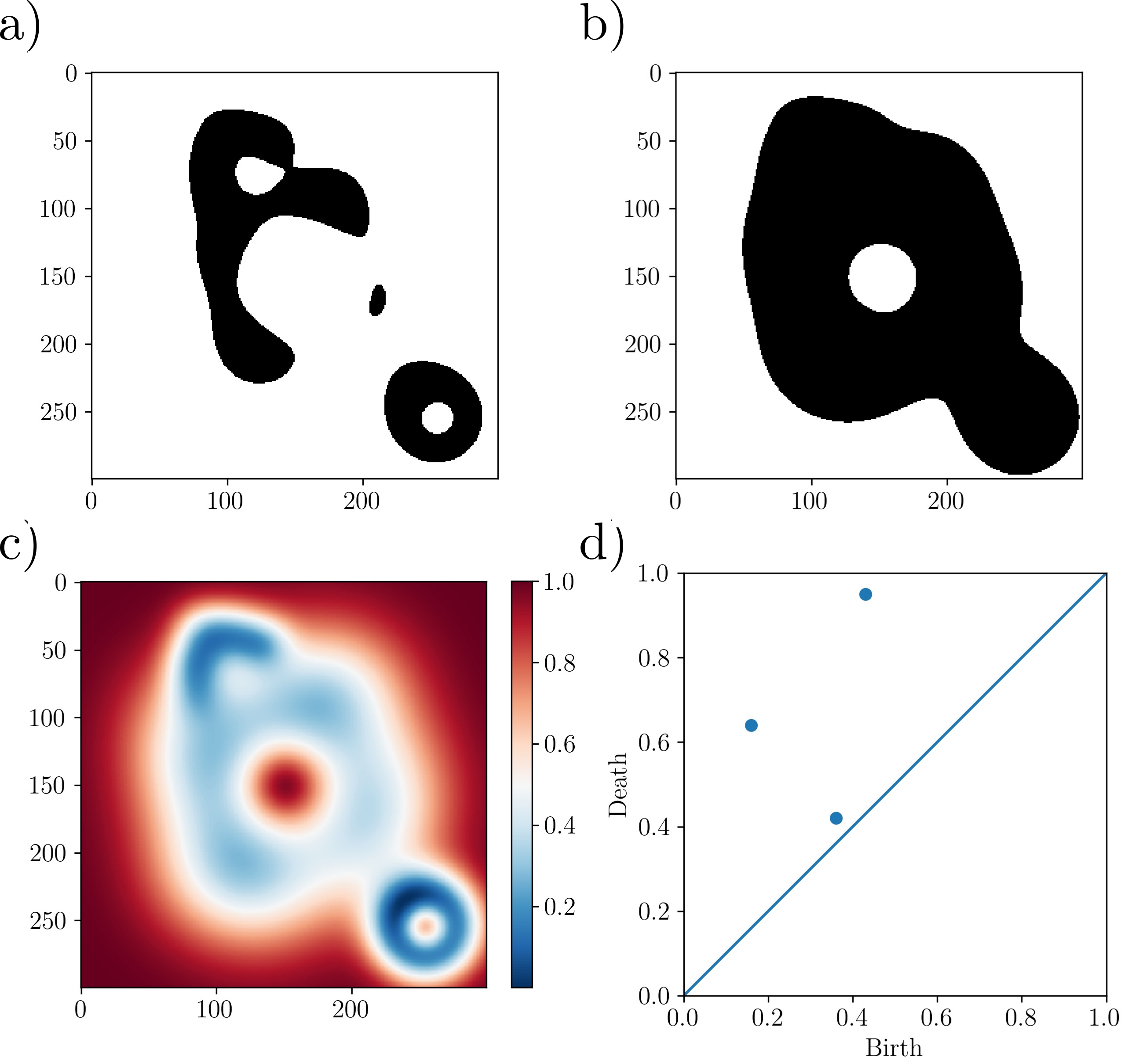}
\caption{a,b) Sublevel sets of the image given in c. d) Persistence diagram of the image shown in c. }
\label{fig:sublevel_set_persistence}
\end{figure}
\begin{figure}[t!]
\centering
\includegraphics[width=0.5\textwidth,keepaspectratio]{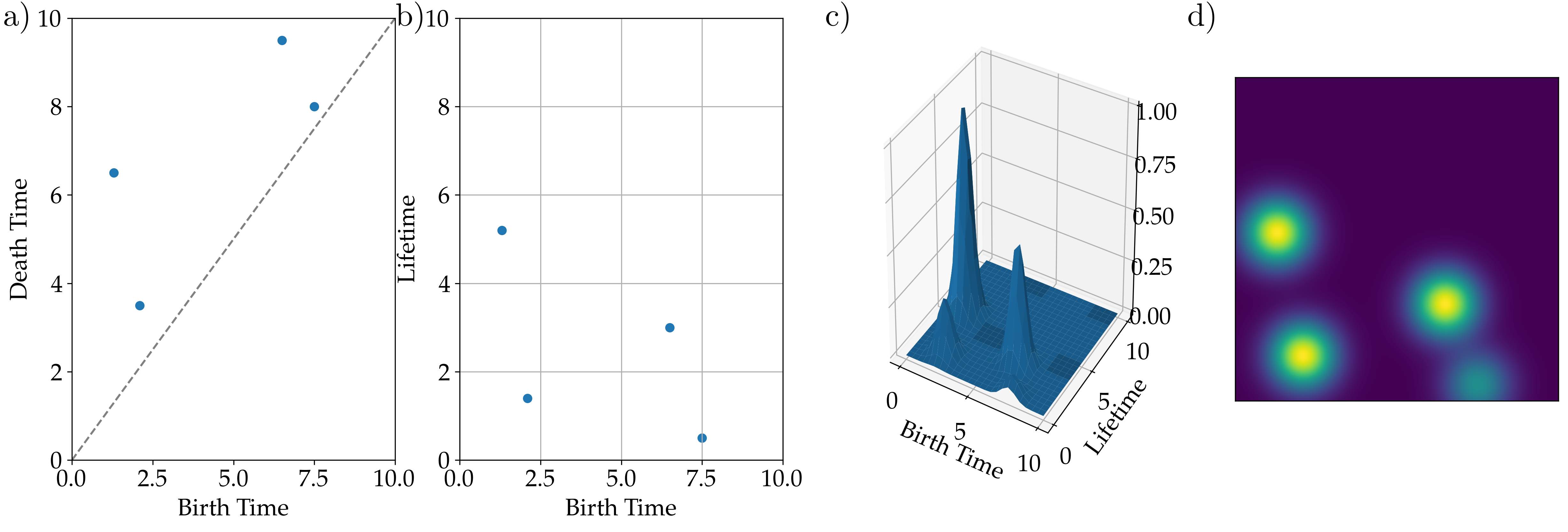}
\caption{The steps to obtain a persistence image.}
\label{fig:per_img}
\end{figure}
In addition to standard signal processing tools used in Secs.~\ref{sec:Gaussian} and~\ref{sec:FFT}, we use persistent homology from TDA to extract features from synthetic surfaces.
Persistent homology is the flagship tool from TDA, and it analyzes the shape of the data. We briefly explain persistent homology in the next section, and refer to Refs.~\cite{Munch2017,Munkres2018,Ghrist2008} for more details. 
\subsubsection{\textbf{Background}}
We use the sublevel sets of the images (see Fig.~\ref{fig:sublevel_set_persistence}a and ~\ref{fig:sublevel_set_persistence}b) and of surface profiles.
Let $f$ be a function that represents our data set such that $f:\mathcal{X} \rightarrow \mathbb{R}$. The domain of surface profiles or surfaces is denoted as $\mathcal{X}$. 
Then, the sublevel sets of $f$ is defined as 
\begin{equation*}
L_{\lambda}= \{x: f(x)\leq \lambda\} = f^{-1}([-\infty,\lambda)),
\end{equation*}
where $\lambda$ is a threshold \cite{Berry2020}. 
The sorted set of threshold values, ${\lambda_{1}<\lambda_{2}<\ldots <\lambda_{l}}$ forms an ordered collection of subsets such that
\begin{equation*}
L_{\lambda_{1}} \subseteq L_{\lambda_{2}}\subseteq \ldots \subseteq L_{\lambda_{l}}.
\end{equation*}
The collection of these ordered sets $\mathcal{L} = \cup_{\lambda}{L_{\lambda}}$ is called filtration with respect to $f$. 

Persistent homology tracks the changes in a given filtration. For instance, persistent homology in 0D is concerned with connected components, while the homology in 1D tracks loops.
In this study, we work with both 0D and 1D persistent homology.   
The threshold value where a topological feature is observed for the first time is called the birth time of that feature. 
When the feature disappears, the corresponding threshold is denoted as the death time of the feature.
For instance, a loop can first appear (is born) at threshold $\lambda_{i}$ and it can fill in (die) at $\lambda_{j}$. 
Pairs of birth and death times for each topological feature are plotted in a persistence diagram (see Fig.~\ref{fig:sublevel_set_persistence}d).

Working directly with persistence diagrams is not easy due to their complex structure and we cannot simply perform algebraic operations with persistence diagrams since the number of topological features can be different for each diagram \cite{Berry2020}. 
Therefore, we extract features from persistence diagrams using their functional summaries. 
We employ three methods to featurize the diagrams, namely Carlsson coordinates \cite{Adcock2016,Khasawneh2018}, persistence images \cite{Adams2017} and template functions \cite{Perea2019}.
   
\subsubsection{\textbf{Carlsson Coordinates}}
Carlsson coordinates are the set of functions of birth and death points of the persistence diagrams. The first four of them are introduced in Ref.~\cite{Adcock2016}, while the last one was added in Ref.~\cite{Khasawneh2018}. Definition of the five coordinates is given as
\begin{gather*}
f_{1}(\mathcal{D}) = \sum b_{i}(d_{i}-b_{i})\\
f_{2}(\mathcal{D}) = \sum (d_{max}-d_{i})(d_{i}-b_{i})\\
f_{3}(\mathcal{D}) = \sum b_{i}^{2}(d_{i}-b_{i})^{4}\\
f_{4}(\mathcal{D}) = \sum (d_{max}-d_{i})^{2}(d_{i}-b_{i})^{4}\\
f_{5}(\mathcal{D}) = max{(d_{i}-b_{i})},
\end{gather*}
where the $b_{i}$ and $d_{i}$ are the birth and death times of the features, $\mathcal{D}$ represents persistence diagram, and $d_{max}$ is the maximum death time in the persistence diagram.  
These five features are computed for each persistence diagram to generate a feature matrix which can be used in supervised classification algorithms to classify surfaces and profiles.
\subsubsection{\textbf{Persistence Images}}
Another featurization technique for persistence diagrams is to convert them into persistence images \cite{Adams2017}. The first step is to apply a linear transformation
\begin{equation*}
T(b_{i},d_{i}) = (b_{i},d_{i}-b_{i})
\end{equation*}
which converts persistence diagram $\mathcal{D} = \{b_{i},d_{i}\}$ to lifetime-birth time diagram (see Fig.~\ref{fig:per_img}b). Then, a normalized Gaussian is defined as 
\begin{equation*}
G_{k}(x,y) = \frac{1}{2\pi \sigma^{2}} e^{[(x-b_{k})^{2}+(y-p_{k})^{2}]/2\sigma^{2}}
\end{equation*}
where $p_{k}$ is the lifetime of the points in the diagram, and $\sigma$ is the standard deviation \cite{Adams2017}. 
$G_{k}(x,y)$ is placed at the points of the transformed diagram. We also define a weighting function
\[W(b_{k},p_{k})= \begin{cases} 
      0 & p_{k}\leq 0 \\
      p_{k}/p_{max} & 0\leq p_{k}\leq p_{max} \\
      1 & p_{max}\leq p_{k} 
   \end{cases}.
\]
Then, the persistence surface shown in Fig.~\ref{fig:per_img}c is defined using the weighting function and the Gaussians such that 
\begin{equation*}
S(x,y) = \sum_{(b_{k},p_{k})\in \mathcal{D}} W(b_{k},p_{k})D_{k}(x,y).
\end{equation*}
A grid can be defined over the domain of the persistence surface $S$ with respect to the chosen pixel size and range of birth time and death time. 
The persistence surface is integrable over the grid, and pixel values of persistence images (see Fig.~\ref{fig:per_img}d) is found by
\begin{equation*}
I_{i,j}(S)=\iint Sdxdy.
\end{equation*}

The feature vector for a persistence image is obtained by concatenating the rows of the persistence image, and can then be input to supervised machine learning algorithm.

\subsubsection{\textbf{Template Functions}}
Template functions were introduced in Ref.~\cite{Perea2019}. Given a persistence diagram $\mathcal{D}$, we convert its coordinate system into birth-lifetime diagram.
%All the points in $\mathcal{D}$ are in the upper half plane $\mathbb{W}:=\{(x,y)\in \mathbb{R}^{2}: 0\leq x<y \}$\cite{Perea2019}.

A template function for a persistence diagram is defined as 
\begin{equation*}
v_{f}(\mathcal{D}) = \sum_{(b,p) \in \mathcal{D}} f(b,p),
\end{equation*}
where $b$ and $p$ represent the birth time and lifetime, respectively. The set of template functions forms a template system $\mathcal{T}$. For more details about template functions and template systems, one can refer to Ref.~\cite{Perea2019}. Here we define a template system of Chebyshev polynomials using
\begin{equation*}
f(x,y) = \beta (x,y)\cdot | \ell_{i}^\mathcal{A}(x) \ell_{j}^\mathcal{B}(y) |,
\end{equation*}
where $l_{i}^{\mathcal{A}}$ and $l_{i}^{\mathcal{B}}$ are the Lagrangian functions \cite{Perea2019} computed on mesh $\mathcal{A}$ and $\mathcal{B}$ which are defined to include all points in the persistence diagram. 

%% file: sections/sec-results_and_discussion.tex
%!TEX root = ../ICMLA_2021.tex
%*****************************
\section{Results and Discussion}
\label{sec:Results}
This section compares the results obtained from feature extraction methods explained in Sec.~\ref{sec:method} and includes our concluding remarks.
We apply 10-fold cross validation while training and testing the performance of four supervised classification algorithms: 
support vector machine (SVM), logistic regression (LR), random forest (RF) and gradient boosting (GB).
In order to be consistent, the same random state number for cross validation is used to generate the same sets for training and testing for all feature extraction methods.

\begin{figure}[h]
\centering
\includegraphics[width=0.5\textwidth,keepaspectratio]{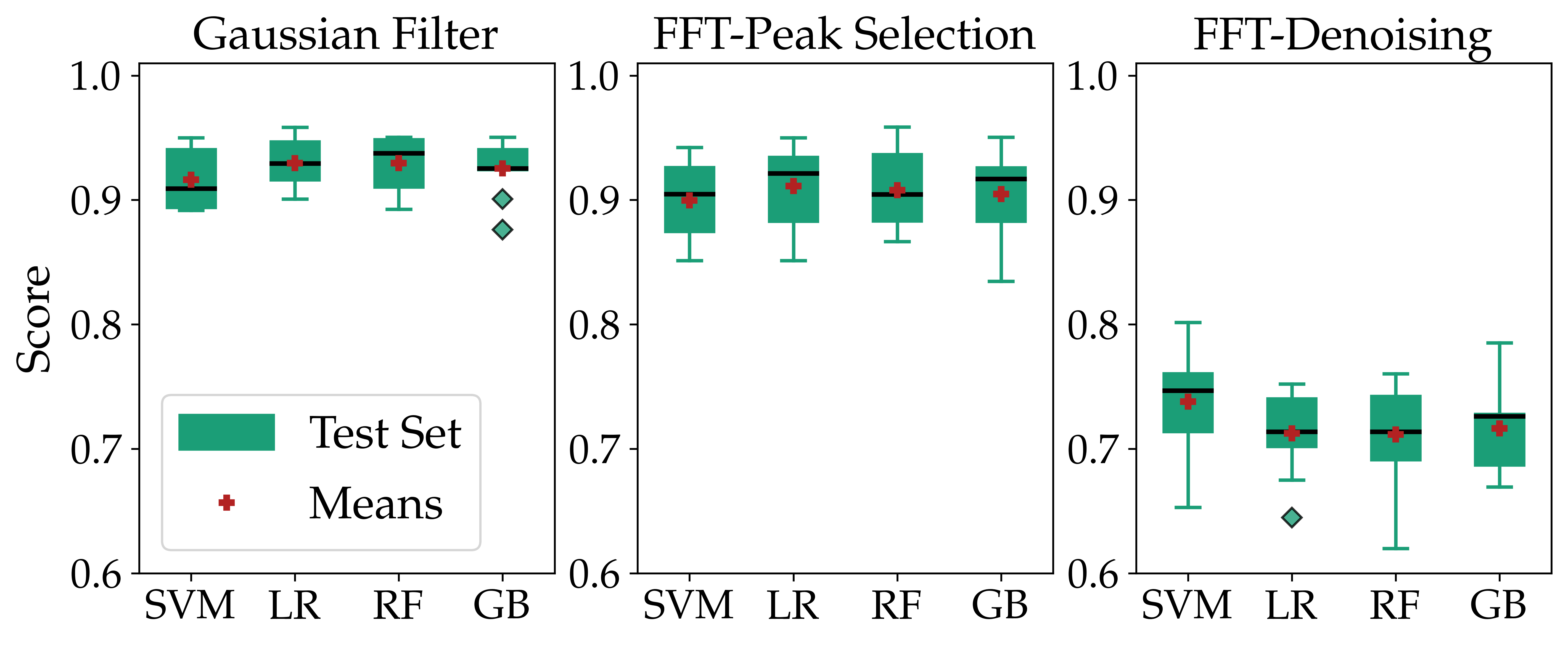}
\caption{Surface profile test set results obtained with 1D implementation of traditional signal processing tools.}
\label{fig:1D_Traditional_Results}
\end{figure}

We compare the performance of each feature extraction method with respect to accuracy of classification using default parameters for all classification algorithms. 
The resulting accuracies of each classifier is represented with box plots.
Figure~\ref{fig:1D_Traditional_Results} shows surface profile classification results obtained from traditional signal processing approaches. 
It is seen that Gaussian Smoothing and peak selection from FFT and PSD plots outperform the FFT method where we denoise the signal by setting a thereshold in its spectrum.
There is no significant difference between the results of four classifiers as seen from the figure.
%However, we observe overfitting in random forest and gradient boosting as shown in middle plot of Fig.~\ref{fig:1D_Traditional_Results}.
%This overfitting can be resolved by tuning the parameters of these classifiers, however parameter tuning is not in the scope of this work.
% Parameter tuning can help us to improve the classification accuracies, however parameter tuning is not in the scope of this work.
%%%% Figure %%%%%
%%%% ----- %%%%%

\begin{figure}[h]
\centering
\includegraphics[width=0.5\textwidth,keepaspectratio]{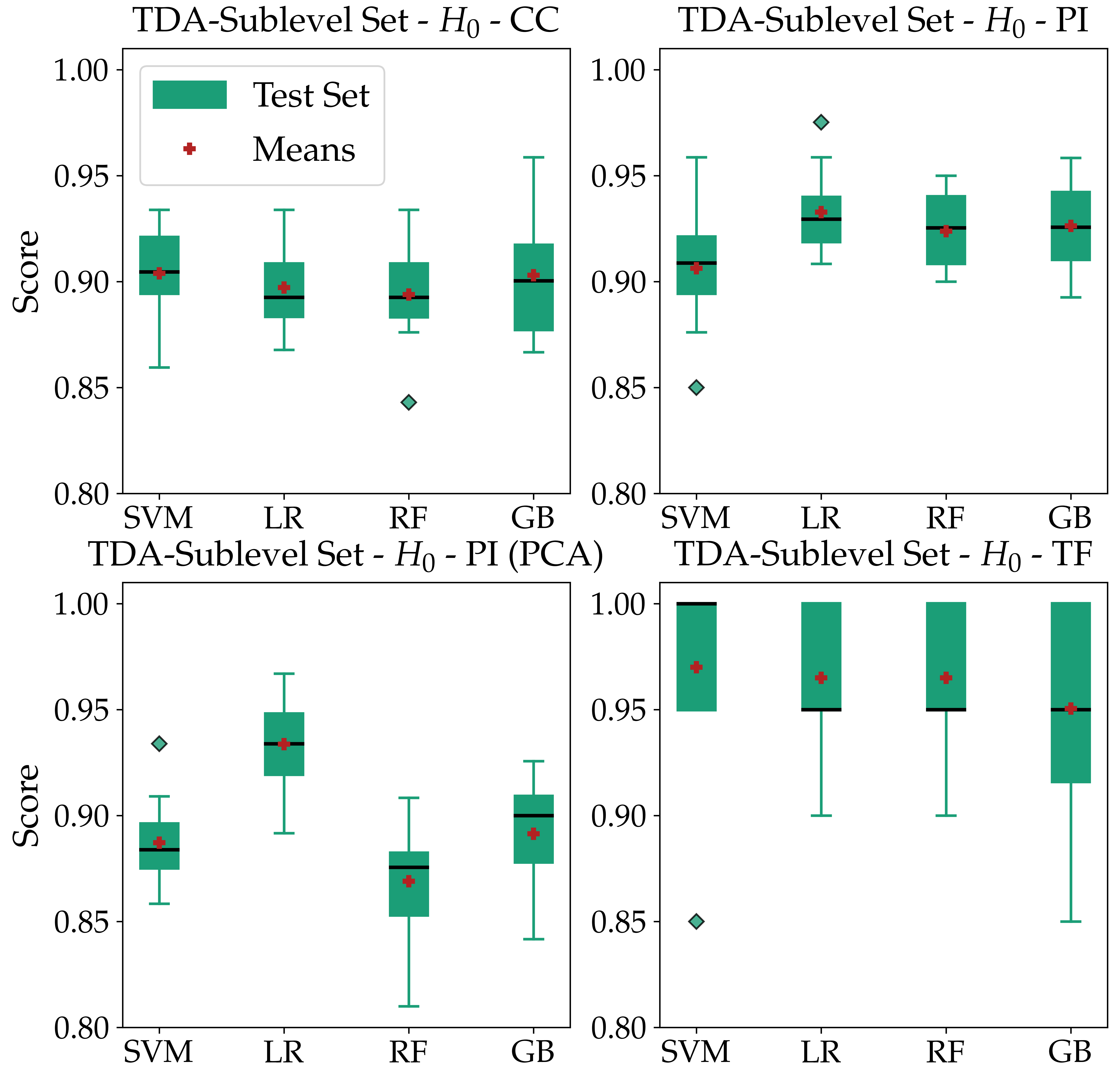}
\caption{Surface profile classification results obtained with OD ($H_{0}$) sublevel set persistence. Carlsson Coordinates (CC), persistence images (PI) and template functions (TF) are used to extract features. First plot in second row represents the results after we apply dimensionality reduction to features obtained from persistence images.}
\label{fig:0D_TDA_Results}
\end{figure}

The second approach we used to extract features from surface profiles is the TDA based approach. 
0D sublevel set persistence is utilized to compute persistence diagrams, and feature extraction is performed using the three methods explained in Sec.~\ref{sec:TDA-based}.
Figure~\ref{fig:0D_TDA_Results} provides the resulting test set accuracies for four classifiers.
We also applied Principal Component Analysis (PCA) to the features obtained from persistence images.
The first 10 components with the highest variance ratios are used to project the feature space onto a 10 dimensional space. 
The resulting feature matrix is used for classification and the corresponding accuracies are provided in Fig.~\ref{fig:0D_TDA_Results}.
It is seen that all three feature extraction methods give mean accuracies greater than or around 90\%.
We notice that this dimensionality reduction does not increase the accuracy of the classifiers for persistence images.
The chosen 10 components may not correspond to regions where a Gaussian is placed, so we may accidentally eliminate an important descriptor while reducing the dimension of the feature space.
For surface profile data, the dimension of the feature space is reduced from 320 to 10. 
This provides an advantage in terms of the time required for classification, and the resulting mean accuracies are still around 90\%. Thus, it is worthwhile to apply PCA on persistence image features.
%\twocolumn[{\includegraphics[width=1\textwidth,keepaspectratio]{Box_Plot_TDA_Profiles_H0_PT0_cv_10.png}}]

\begin{figure}[h]
\centering
\includegraphics[width=0.5\textwidth,keepaspectratio]{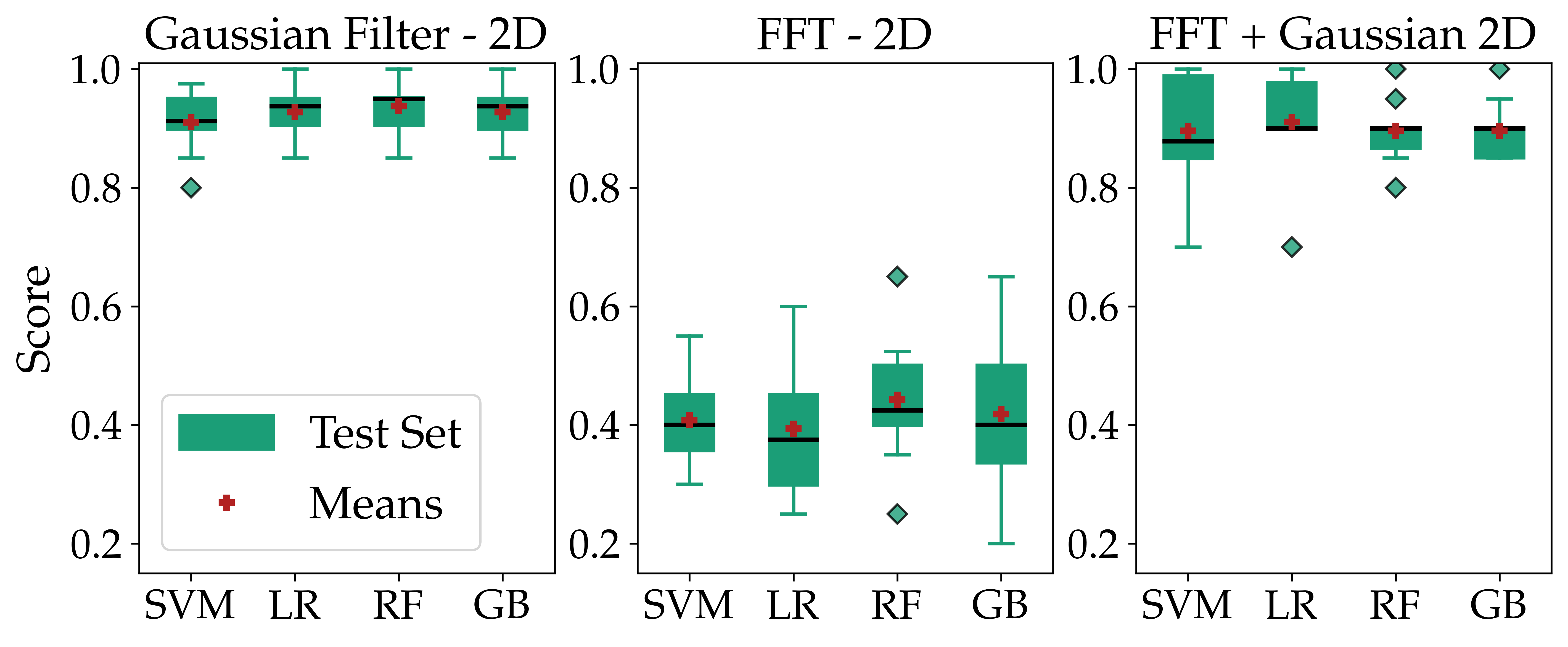}
\caption{Results of surface classification obtained with 2D implementation of signal processing tools.}
\label{fig:2D_Traditional_Results}
\end{figure}

\begin{figure}[h]
\centering
\includegraphics[width=0.5\textwidth,keepaspectratio]{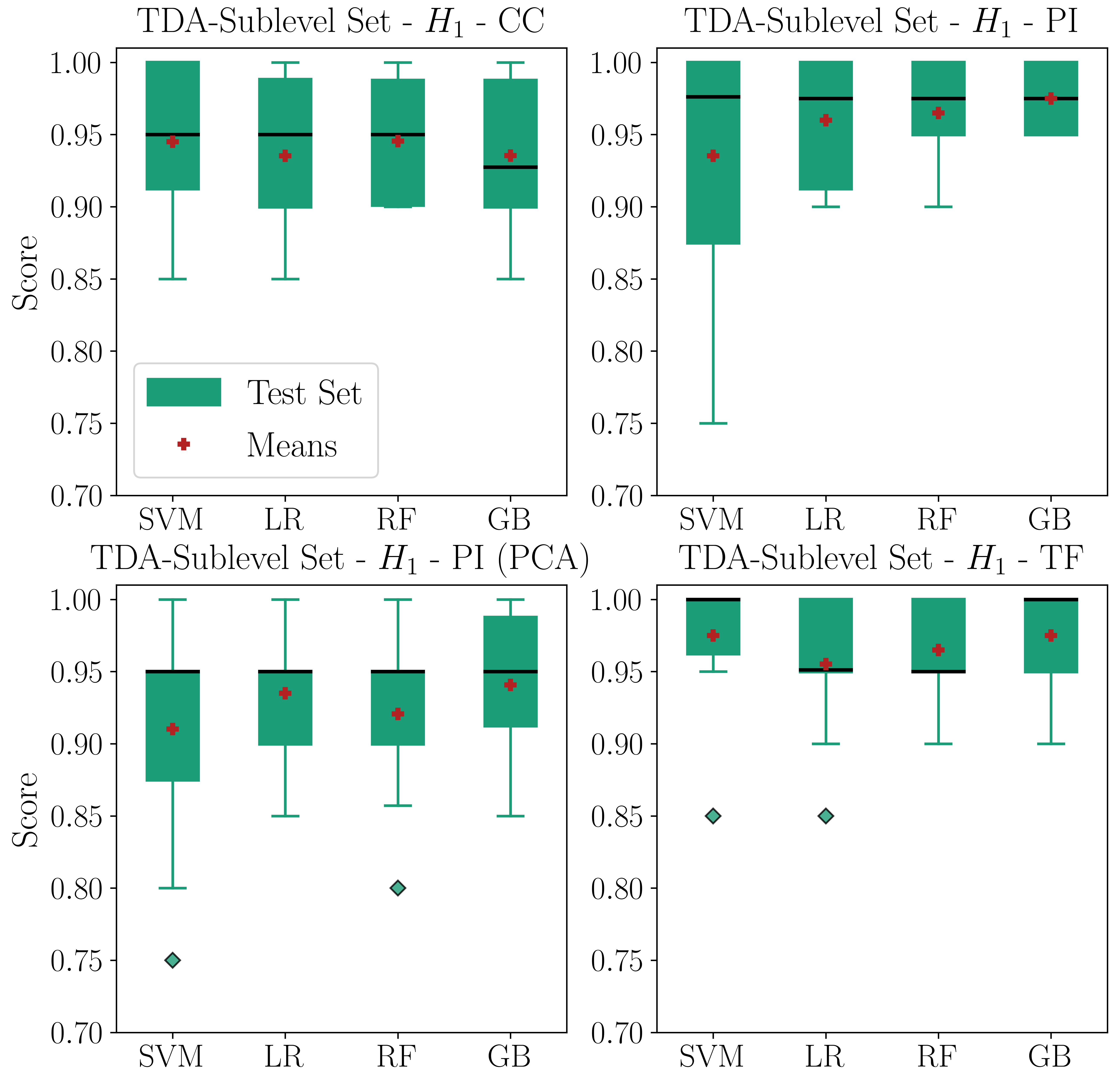}
\caption{Results of surface classification obtained with 1D persistence $H_{1}$ using Carlsson coordinates (CC), persistence images (PI) and template functions (TF).}
\label{fig:2D_TDA_Surface_Results}
\end{figure}

Surface classification results for traditional signal processing tools are provided in Fig.~\ref{fig:2D_Traditional_Results}.
It is seen that Gaussian smoothing combined with FFT provide the highest scores.
However, directly applying FFT in 2D on surface measurements yields poor results for all classifiers.
This shows the importance of obtaining the roughness component of a given surface.
%In combination of FFT and Gaussian smoothing, the reference surface is found by Gaussian filtering and then subtracted from the main surface to obtain roughness surface.
%FFT in 2D is then applied on the roughness component and this improves the classification accuracy significantly as shown in Fig.~\ref{fig:2D_Traditional_Results}.
In addition, we provide the results obtained with TDA based approach for surface classification in Fig.~\ref{fig:2D_TDA_Surface_Results}.
All feature extraction methods from persistence diagrams yield mean accuracies above 90\%.
We also apply PCA to the feature space of persistence images as we classify surface profiles.
Again, application of PCA does not increase the classification accuracy, but it still provides mean accuracies around 95\%. 
0D ($H_{0}$) and 1D ($H_{1}$) persistence provide similar results, so we only provide 1D persistence ($H_{1}$) results in Fig.~\ref{fig:2D_TDA_Surface_Results}.
The highest mean accuracies are obtained by using the template function methods for both of them.

Results of traditional and TDA-based approach for profile/surface classification are comparable. However, the TDA-based approach provides three main advantages: 1) it requires no parameter selection, 2) it provides an automatic and systematic way for feature extraction, and 3) it allows adaptive feature extraction.
Traditional methods do not share all these advantages. For instance, we need to select two restriction parameters (MPD and MPH) and a kernel size for FFT/PSD and Gaussian 2D implementations, respectively. In addition, the selection of the MPD and MPH or kernel size for Gaussian smoothing require visually inspecting the spectra thus they have low automation potential.
These parameters and thresholds are typically selected using only a small portion of the data set.
The selected parameters may not be suitable for every surface profile or surface, so traditional signal processing approaches are non-adaptive. 

In this study, we proposed an automatic and adaptive feature extraction approach to determine the level of roughness in profile and areal measurements of surfaces.
The proposed approach yields similar classification accuracies for surface and profile classification, and it eliminates the manual preprocessing which is required by traditional signal processing tools. 
All of the results in this study are obtained using the default parameters of the classification algorithms. 
Further parameter tuning for each classifiers can result in better classification accuracies with smaller deviations. 
In addition, TDA-based approaches are computationally expensive compared to highly optimized traditional signal processing algorithms. 
Therefore, future work of this study will include parameter tuning and optimizing TDA-based approaches to speed up the computation, as well as applying the proposed approach to experimental data.